\documentclass[fleqn,usenatbib]{mnras}

\usepackage[T1]{fontenc}

\DeclareRobustCommand{\VAN}[3]{#2}
\let\VANthebibliography\thebibliography
\def\thebibliography{\DeclareRobustCommand{\VAN}[3]{##3}\VANthebibliography}

\usepackage{graphicx}	
\usepackage{amsmath}	
\usepackage{amssymb}	
\usepackage{newtxtext,newtxmath}
\usepackage{orcidlink}

\title[Measuring neutron-star distances and properties with gravitational-wave parallax]{Measuring neutron-star distances and properties with gravitational-wave parallax}

\author[M. Sieniawska, D. I. Jones, A. L. Miller]{
Magdalena Sieniawska,$^{1}$\thanks{E-mail: magdalena.sieniawska@uclouvain.be}
David Ian Jones,$^{2}$\thanks{E-mail: D.I.Jones@soton.ac.uk} Andrew L. Miller\orcidlink{0000-0002-4890-7627} $^{1,3,4}$\thanks{E-mail: andrew.miller@nikhef.nl}
\\
$^{1}$Centre for Cosmology, Particle Physics and Phenomenology (CP3), Universit\'{e} catholique de Louvain, Chemin du Cyclotron 2, B-1348 Louvain-la-Neuve, Belgium \\
$^{2}$Mathematical Sciences and STAG Research Centre, University of Southampton, Southampton SO17 1BJ, United Kingdom \\
$^3$Nikhef -- National Institute for Subatomic Physics,
Science Park 105, 1098 XG Amsterdam, The Netherlands \\
$^4$Institute for Gravitational and Subatomic Physics (GRASP),
Utrecht University, Princetonplein 1, 3584 CC Utrecht, The Netherlands
}

\date{Accepted XXX. Received YYY; in original form ZZZ}

\pubyear{2023}

\begin{document}
\label{firstpage}
\pagerange{\pageref{firstpage}--\pageref{lastpage}}
\maketitle

\begin{abstract}
Gravitational-wave astronomy allows us to study objects and events invisible to electromagnetic waves. So far, only signals triggered by coalescing binaries have been detected. However, as the interferometers' sensitivities improve over time, we expect to observe weaker signals in the future, e.g. emission of continuous gravitational waves from spinning, isolated neutron stars. Parallax is a well-known method, widely used in electromagnetic astronomical observations, to estimate the distance to a source. In this work, we consider the application of the parallax method to gravitational-wave searches and explore possible distance estimation errors.  We show that detection of parallax in the signal from a spinning down source can constrain the neutron star moment of inertia. For instance, we found that the relative error of the moment of inertia estimation is smaller than $10\%$ for all sources closer than 300 pc, for the assumed birth frequency of 700 Hz, ellipticity $\geq 10^{-7}$ and for two years of observations by the Einstein Telescope, assuming spin down due purely to quadrupolar gravitational  radiation.
\end{abstract}



\section{Introduction}
\label{sec:intro}
Gravitational-wave (GW) astronomy is one of the youngest and most dynamically-growing fields in modern astrophysics. In 2015, the GW era began with the first detection of the binary black-hole (BH) system GW150914 \citep{Abbott2016}. Since then, tens of the compact binary coalescences--double BHs or double neutron stars (NSs)-- have been seen \citep{Abbott2021a, Abbott2021b} by the LIGO \citep{Aasi2015}, Virgo \citep{Acernese2014} and KAGRA \citep{KAGRA_new} Collaboration (LVK). Such systems are called `standard sirens' -- the GW analog of an astronomical standard candle. This means that the distance to the coalescing binary systems depends only on measurable parameters, such as the GW amplitude, frequency and frequency derivative \citep{Schutz1986}.  Additionally, if electromagnetic counterparts can also be seen with conventional telescopes (in the case of double NS mergers), multi-messenger analysis allows tests of cosmological theories, e.g. the measurement of the Hubble constant. Such an analysis \citep{Abbott2017a} was performed for the first multi-messenger detection, the binary NS merger GW170817 \citep{Abbott2017b, Abbott2017c, Abbott2017d, Abbott2021c}. 

So far, only GW signals emitted by coalescing binaries have been detected in the LVK data. However, as the network of GW observatories grows, the sensitivity of the instruments is improving and data analysis methods are constantly progressing; thus, weaker signals are expected to be detected in the future, in particular those emitted by isolated, rotating and asymmetric NSs. These signals would be long-lasting and almost monochromatic. Such so-called `continuous gravitational waves' (CGWs) could be triggered by the rigid rotation of a triaxial star, whose deformation is supported by elastic and/or magnetic strains (see \citealt{Andersson2011, Lasky2015, Riles2017, Sieniawska2019, Piccinni2022} for reviews). Searches for CGW signals were performed in the past in LVK data (for example, see \citealt{Abbott2017e, Abbott2018, Abbott2019, Abbott2020a, Abbott2021, Abbott2022}). Even though no detections were claimed, astrophysically interesting upper limits
were set.

As was shown in \citet{Sieniawska2022}, CGW signals cannot be used as canonical standard sirens with the search analysis methods used so far by LVK, since their distances are always degenerate by one of the unknown physical parameters: the ellipticity (which measures a star's level of deformation) or the moment of inertia of the NS. Here, we propose a way to break this degeneracy by using GW parallax. GW parallax is analogous to the astronomical parallax, broadly used in space astrometry and distance determination to relatively nearby objects, e.g. by the Gaia mission \citep{Gaia2016a,Gaia2016b}. Parallax (both GW and electromagnetic) is the apparent displacement of the position of the source against the background of distant objects, caused by the change of the observer's point of view over time.

A theoretical consideration of the GW parallax was presented in \citet{Seto2005}. Here, we give an updated and simplified analysis of this problem. The paper is organized as follows: in Sect.~\ref{sec:method} we provide general information about gravitational radiation theory, our signal model, CGW detectability and the GW parallax method. We also estimate distance and moment of inertia errors. Section~\ref{sec:results} contains results of our simulations. In Sect.~\ref{sec:discussion} we present some discussion, while in Sect.~\ref{sec:conclusions} we conclude our work.

\section{Method}
\label{sec:method}
For CGW emission from a spinning NS, approximated as a triaxial ellipsoid, the amplitude of the signal $h_0$ can be parameterised as \citep{Ostriker1969, Melosh1969, Chau1970, Press1972, Zimmermann1978}:
\begin{equation}
    h_0 = \frac{4G}{c^4}\frac{1}{d}I_3\epsilon\omega_{\rm rot}^2,
    \label{eq:h0tr}
\end{equation}
where $\epsilon$ is the ellipticity that measures a star's deviation from a spherical shape, defined as $\epsilon \equiv (I_2 - I_1)/I_3$, where $I_3$ is the moment of inertia about the spin axis, with $I_1$ and $I_2$  moments of inertia along axes perpendicular to $I_3$, $\omega_{\rm rot} = 2 \pi f_{\rm rot}$ is the (angular) rotational frequency, and $d$ the distance to the source. For the triaxial ellipsoid model, $f_0 = 2 f_{\rm rot}$, where $f_0$ is the CGW frequency and $f_{\rm rot}$ is the rotational frequency.

As was mentioned in Sect.~\ref{sec:intro}, a method to use a CGW detection to measure the combination $d/\sqrt{I_3}$ was recently proposed by \citet{Sieniawska2022}.  In this work we propose to use GW parallax to provide an independent measurement of distance $d$ to the source. Then, by combining $d/\sqrt{I_3}$ and GW parallax measurements, one can break the degeneracy between $d$ and $I_3$ and thereby measure $I_3$.

The angular parallax $\delta\theta$ is related to the distance to the source $d$ as:
\begin{equation}
    \delta\theta =\frac{R_{\rm orb}}{d} = 10^{-3} {\, \rm arcsec \,} \left(\frac{1 \, \rm kpc}{d}\right) ,
    \label{eq:parallax_general}
\end{equation}
where $R_{\rm orb}$ is the distance from the Earth to the Sun, and we work in the small angle limit where $R_{\rm orb} \ll d$.  It follows that our ability to observe parallax in a CGW signal is determined by our ability to resolve the sky location of a source.  We will therefore proceed to make a rough estimate of the angular resolution of a CGW observation.

\begin{figure}
	\includegraphics[width=\columnwidth]{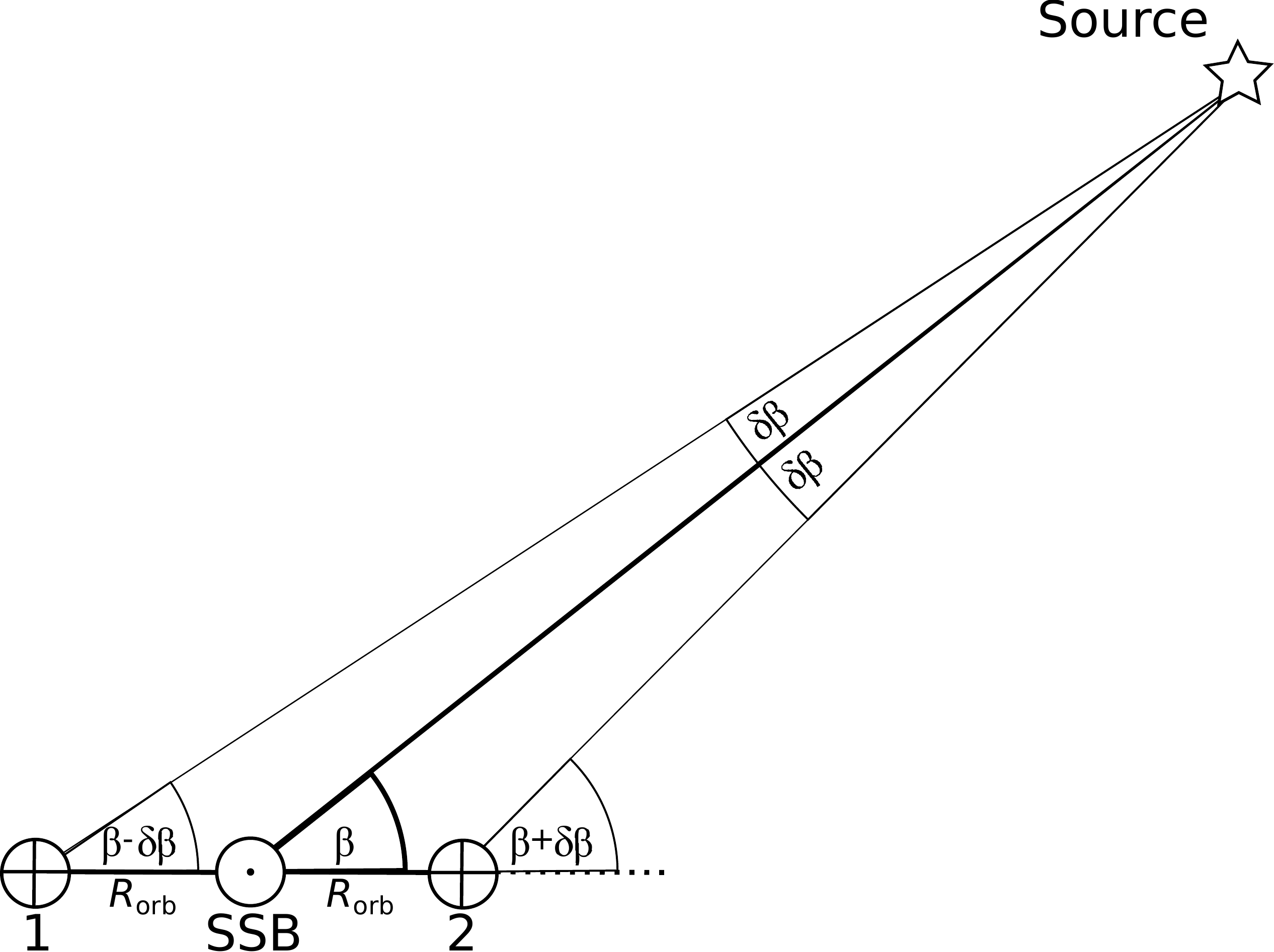}
    \caption{Schematic explanation of the GW parallax effect, reduced to the 2-dimensional plane. The angle $\beta$ is the (fixed) ecliptic latitude of the CGW source, while $\delta \beta$ is the \emph{apparent} shift in its location due to the Earth's orbital motion.}
    \label{fig:schem}
\end{figure}

The sky resolution of CGW observation comes mainly from the sky direction-dependent Doppler shift that the Earth's motion about the Sun induces on the signal \citep{Jaranowski1999}.  We approximate the movement of the detector with respect to the Solar System Barycenter (SSB) to be circular with radius $R_{\rm orb}=1$ astronomical unit [AU], the distance from Earth to the Sun. Then the frequency shifts due to the Doppler effect on a signal of frequency $f_0$ can be approximated as
\begin{equation}
\Delta f = \frac{\Omega_{\rm orb}R_{\rm orb}f_0\cos\beta }{c},
\label{eq:ft}
\end{equation}
where $\Omega_{\rm orb} \approx 2 \cdot 10^{-7}$ [rad/s] is the orbital frequency of the Earth around the Sun and $\beta$ is the ecliptic latitude of the source. In Fig.~\ref{fig:schem}, we present a schematic geometry of the system, reduced to the 2-dimensional projection. In 6 months, the Earth changes its position from 1 to 2.  The apparent location of the source, \emph{relative to the detector}, shifts by an amount $\pm \delta \beta$, with the increase $+\delta \beta$ and decrease $-\delta \beta$  being of equal magnitude in the small angle limit $(R_{\rm orb} / d \ll 1$).

In fact, the CGW analyses commonly carried out employ timing models where the sky location of the source is specified using angular coordinates right ascension and declination, i.e. angles with respect to the (fixed) ecliptic coordinate system.  Of course, the source remains fixed with respect to such coordinates; parallax is only an \emph{apparent} shift in sky location of a foreground source relative to a distant background.  However, currently employed CGW timing models neglect such terms (see discussion in \citealt{Jaranowski1999}, just before their equation (A20)).  Having made such a zero-parallax approximation, parallax will manifest itself as a yearly variation in other parameters, including the apparent values of the sky location, which is why it is appropriate for us to consider the sky resolution of the CGW search when considering the parallax effect.  

Ideally, one would measure the distance by retaining the finite-distance terms, so that distance $d$ explicitly appears in the waveform model, and can them be extracted along with all other source parameters.  In practice, one could adopt a more pragmatic approximate approach, where one divides the data set into  segments of length significantly less than a year (e.g. three months), and analyses each segment separately.  The parallax effect would then manifest itself as an apparent shift in the sky location of the source between segments, with a one year modulation period.  This approximate technique has the advantage of making use of existing wave form models.  

Such a search would have to contend with \emph{three} sorts of frequency variation.  There would be the secular spin-down of the star, caused by its steady loss of kinetic energy as it radiates gravitationally and/or electromagnetically.  There would be an annual modulation due to the Doppler effect, caused by the Earth's motion around the sun.  Superimposed on this would be an additional small annual modulation from the parallax effect.  The first of these can be eliminated by using a sufficiently accurate spin down model (i.e.\ a sufficiently fine template back of the spin down derivatives appearing in a Taylor expansion of $f_0(t)$).  The second can be eliminated by using a sufficiently fine sky-grid when performing the Doppler demodulation.  The third would then appear as an annual modulation that further refinement of the sky grid would \emph{not} remove.  Only the inclusion of the finite distance terms in the waveform model would allows its elimination (thereby giving the distance explicitly as a model parameter).  For observations short compared to one year, the secular variation in frequency will inevitably be degenerate with the annual one.  But for longer observations, the different temporal evolutions of the secular and annual variations will always all for their separation.  Indeed, searches for CGWs routinely apply (separate) frequency corrections to deal with both the secular and annual variations, see e.g. \citet{riles_22}. 

With these measurement issues noted, we will restrict ourselves here to estimating for which CGW signals parallax is potentially detectable, and to what accuracy.

We first begin by estimating the CGW sky resolution.  To do so, we exploit the difference in the magnitude of the Doppler effect $\Delta f$ for two slightly different sky locations, separated by some small angle $\delta \beta$. By calculating the partial derivative of Eq.~\ref{eq:ft}, one can estimate the corresponding variation in the magnitude $\delta f$ of the Doppler modulation:
\begin{equation}
\delta f \approx \left| \frac{\partial \Delta f}{\partial \beta} \right| \delta \beta 
=  \frac{\Omega_{\rm orb}R_{\rm orb}f_0\sin\beta}{c} \delta \beta.
\label{eq:delb}
\end{equation}
It follows that our Doppler-induced sky-resolution $\delta \beta$ is in turn determined by our uncertainty in the CGW frequency, $\delta f$.

We assume that the frequency uncertainty $\delta f$ corresponds to the theoretical (minimal) uncertainties of the frequency measurements, taken from the Fisher Information Matrix from \citet{Jaranowski1999}:
\begin{equation}
    \delta f = \sigma(f) = \frac{10\sqrt{3}}{\pi T_{\rm obs}\rho},
    \label{eq:sigmaf}
\end{equation}
where $T_{\rm obs}$ is the observation time and $\rho$ is the signal-to-noise ratio (SNR).  By then combining Eqs.~\ref{eq:delb} and \ref{eq:sigmaf}, we get:
\begin{equation}
    \delta \beta = \frac{10\sqrt{3}c}{\pi T_{\rm obs}\rho \Omega_{\rm orb}R_{\rm orb}f_0 \sin\beta }.
\label{eq:Delta_beta}
\end{equation}
The quantity $\delta \beta$ has the meaning of the sky resolution for a CGW  signal of a given frequency $f_0$, SNR $\rho$, over an observation of duration $T_{\rm obs}$. 

To properly estimate the error in measuring the distance $d$, one should carry out a full Fisher matrix calculation, including finite distance effects so that $d$ appears as a model parameter.  To instead make use of the simple estimates we have made above, we will use an intuitive approach, as follows.  Given that parallax effects are small, we can apply a linearity argument.  Specifically, in the limit of small angular displacements (in which we work throughout this paper), the fractional error in measuring the distance must be proportional to the angular resolution of the search, i.e.\ proportional to $\delta \beta$.  Recall that the quantity $\delta \theta$ defined by equation (\ref{eq:parallax_general}) is the size of the parallax effect itself.  Clearly, the parallax effect is only resolvable if $\delta \beta \lesssim \delta \theta$, with the effect being ``borderline resolvable'' when $\delta \beta \approx \delta \theta$.  By our linearity assumption, it then follows that the fractional accuracy to which the distance can be measured is proportional to $\delta \beta / \delta \theta$.  We will therefore estimate the fractional error in $d$ using this linearity, assuming (in the absence of a full Fisher matrix calculation) a proportionality coefficient of unity:
\begin{equation}
\frac{\sigma(d)}{d} = \frac{\delta \beta}{\delta \theta} .
\end{equation}
Substituting using Eqs.~\ref{eq:parallax_general} and \ref{eq:Delta_beta}, we obtain:
\begin{equation}
\frac{\sigma(d)}{d} = \frac{10\sqrt{3}cd}{\pi T_{\rm obs}\rho \Omega_{\rm orb}R_{\rm orb}^2 f_0 \sin\beta }.
\label{eq:d_rel_err}
\end{equation}
Note that it is only meaningful to use the parallax method in cases where the apparent shift in sky location of the CGW signal is resolvable, i.e. in cases where $\delta \beta \lesssim \delta \theta$, so that $\sigma(d) / d \lesssim 1$.

We checked that the difference between our simple calculation of sky resolution and calculations from \citet{Seto2005} is smaller than one order of magnitude for 1 year $< T_{\rm obs} <$ 10 years. As our model is simpler than the one presented in \citet{Seto2005} and does not include all of the unknown parameters (e.g. the proper motion of the source), the estimated errors presented in this work are typically smaller than the ones in \citet{Seto2005}. We decided to ignore proper motion terms, as, according to \citet{Covas2021}, CGW searches below 1 kHz and $T_{\rm obs} < 1-2$ years cannot measure proper motion.

Having outlined what is required to measure the distance $d$ to a CGW source, and made an estimate of its uncertainty $\sigma(d)$, we now return to the degeneracy issue.  As explained in detail in \citet{Sieniawska2022}, one can extract the combination $d/\sqrt{I_3}$ from a CGW observation of a spinning-down neutron star.  Note, however, the conditions for doing so require that the energy emission from the star be dominated by GW emission, with no other significant energy losses, so that the spin-down can be accurately modelled using standard quadrupolar GW emission.

If this is the case, such that both $d$ (from parallax) and $d/\sqrt{I_3}$ (from spin-down) can be measured, then the degeneracy can be broken, with $I_3$ then being given by the trivial identity
\begin{equation}
    I_3 = \left( \frac{d}{d/\sqrt{I_3}}\right)^2,
    \label{eq:I3}
\end{equation}
with both the numerator $d$ and the denominator $d/\sqrt{I_3}$ being measured quantities. 

We can also estimate the error in such a calculation of $I_3$.  The expression for the distance measurement error obtained with the GW parallax method was given in Eq.~\ref{eq:d_rel_err}.  The error in the quantity $d/\sqrt{I_3}$ was estimated in \citet{Sieniawska2022} to be
\begin{equation}
\begin{aligned}
&\frac{\sigma\left(d/\sqrt{I_3}\right)}{d/\sqrt{I_3}} = &&\\ &=\frac{1}{\rho \pi h_0 d/\sqrt{I_3}}\sqrt{\frac{5G}{4c^3}\biggl[ \frac{75\dot{f}}{f_0^3T_{\rm obs}^2}+  \frac{1620}{f_0 \dot{f}T_{\rm obs}^4 } +  \frac{\pi^2 \dot{f}}{f_0} + \frac{675}{f_0^2T_{\rm obs}^3}  \biggr]}, &&
\label{eq:err_dI3}
\end{aligned}
\end{equation}
where $\dot f$ is the time derivative of the GW frequency.  We can then combine Eqs.~\ref{eq:d_rel_err} and \ref{eq:err_dI3} to evaluate the fractional error in $I_3$, using
\begin{equation}
\left(\frac{\sigma (I_3)}{I_3}\right)^2 = 4 \left[\left( \frac{\sigma(d)}{d} \right)^2 + \left( \frac{\sigma (d/\sqrt{I_3})}{d/\sqrt{I_3}} \right)^2\right].
\label{eq:I3_err}
\end{equation}
Note that this simple error combination ensures that the errors in the two relevant quantities can be treated with the uninformative approach, in which we assume that we do not know how to quantify and take into account the correlations between the statistics, as presented in \citet{Dreissigacker2018}. If in reality both are estimates from one and the same CGW observation, this assumption is of doubtful validity, but a more sophisticated analysis lies beyond the scope of our current study.  We merely flag this as an assumption.

In the next section we will use the above formulae to explore the power of the proposed methods.  To do so, we will need to calculate $\rho$, the SNR.  We do so using the definition of SNR that appears in \citet{Moore2015}:
\begin{equation}
\rho^2 = \int\displaylimits_{f_0}^{f_{\rm end}} \left(\frac{h_c(f)}{h_n(f)}\right)^2 d(\ln f),
\label{eq:snr}
\end{equation}
where $f_0$ is now to be understood as the CGW frequency at the beginning of the observation, and $f_{\rm end}$ the frequency at the end. The quantity $h_{\rm c}(f)$ is characteristic amplitude, defined as:
\begin{equation}
h_{\rm c}(f) = 2f\cdot \vert \tilde{h}(f) \vert,
\label{eq:hc}
\end{equation}
where $\tilde{h}(f)$ is a Fourier transform of the CGW signal \citep{Finn1993}. The above equation can be averaged over sky location and source orientation \citep{Jaranowski1998}, resulting in the averaged characteristic amplitude $\langle h_{\rm c} (f) \rangle = \frac{2}{5} h_{\rm c} (f)$.  The quantity $h_{\rm n}$ is  an effective noise of the detector given by:
\begin{equation}
h_{\rm n}(f) = \sqrt{f\cdot S_h(f)},
\end{equation}
where $S_h$ is the amplitude spectral density (a measure of the sensitivity of the detector).

\section{Results}
\label{sec:results}

\begin{figure}
	\includegraphics[width=\columnwidth]{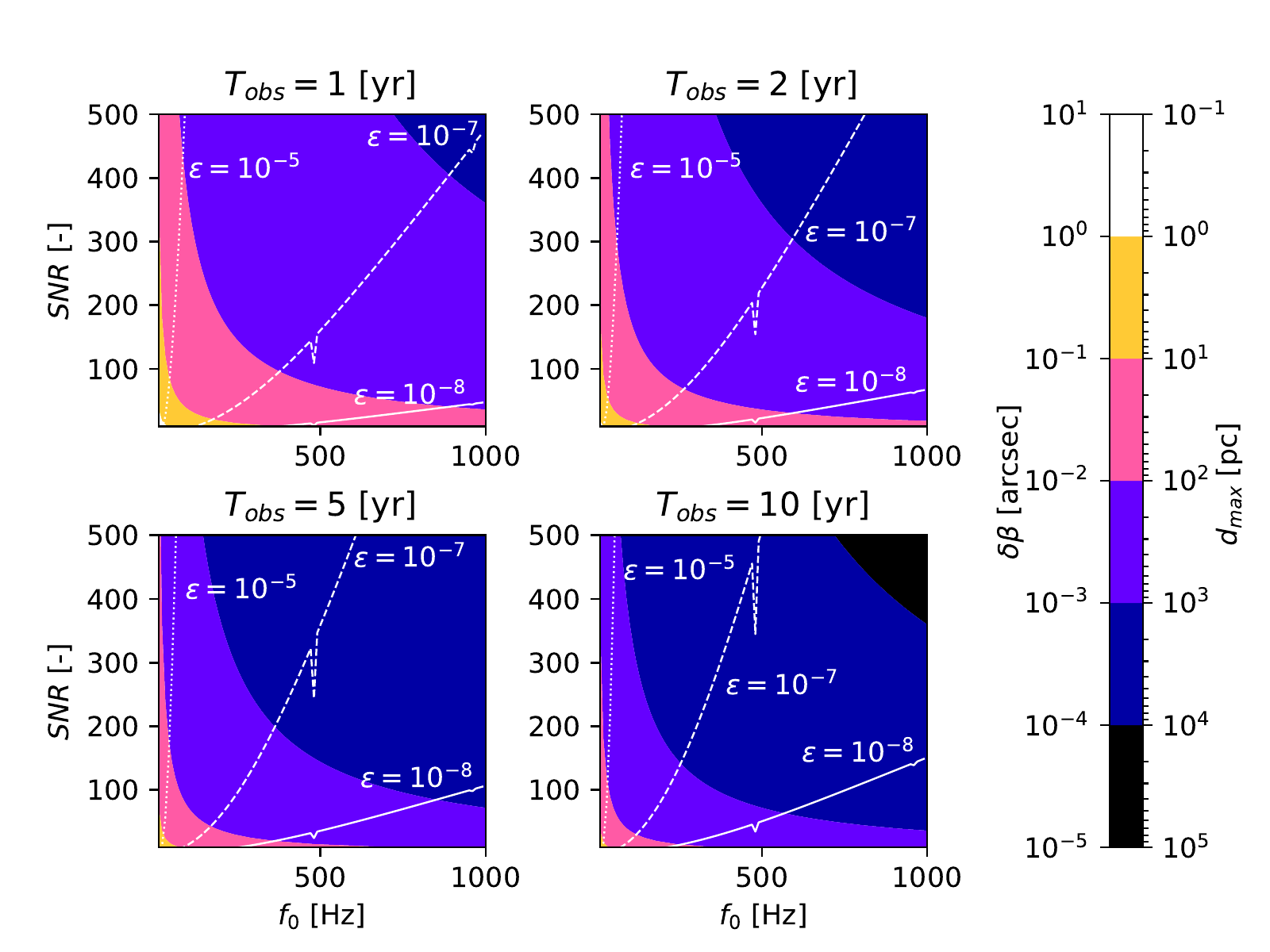}
    \caption{Sky resolution $\delta \beta$ for a broad range of parameters ($f_0, {\rm SNR}, T_{\rm obs}$), together with the corresponding maximal distances $d_{\rm max}$ to which this parallax can be detected. White lines are signal-to-noise ratios calculated for the reference source at 130 pc, assuming the spin-down limit, and for the aLIGO sensitivity.}
    \label{fig:DB_aLIGO}
\end{figure}

\begin{figure}
	\includegraphics[width=\columnwidth]{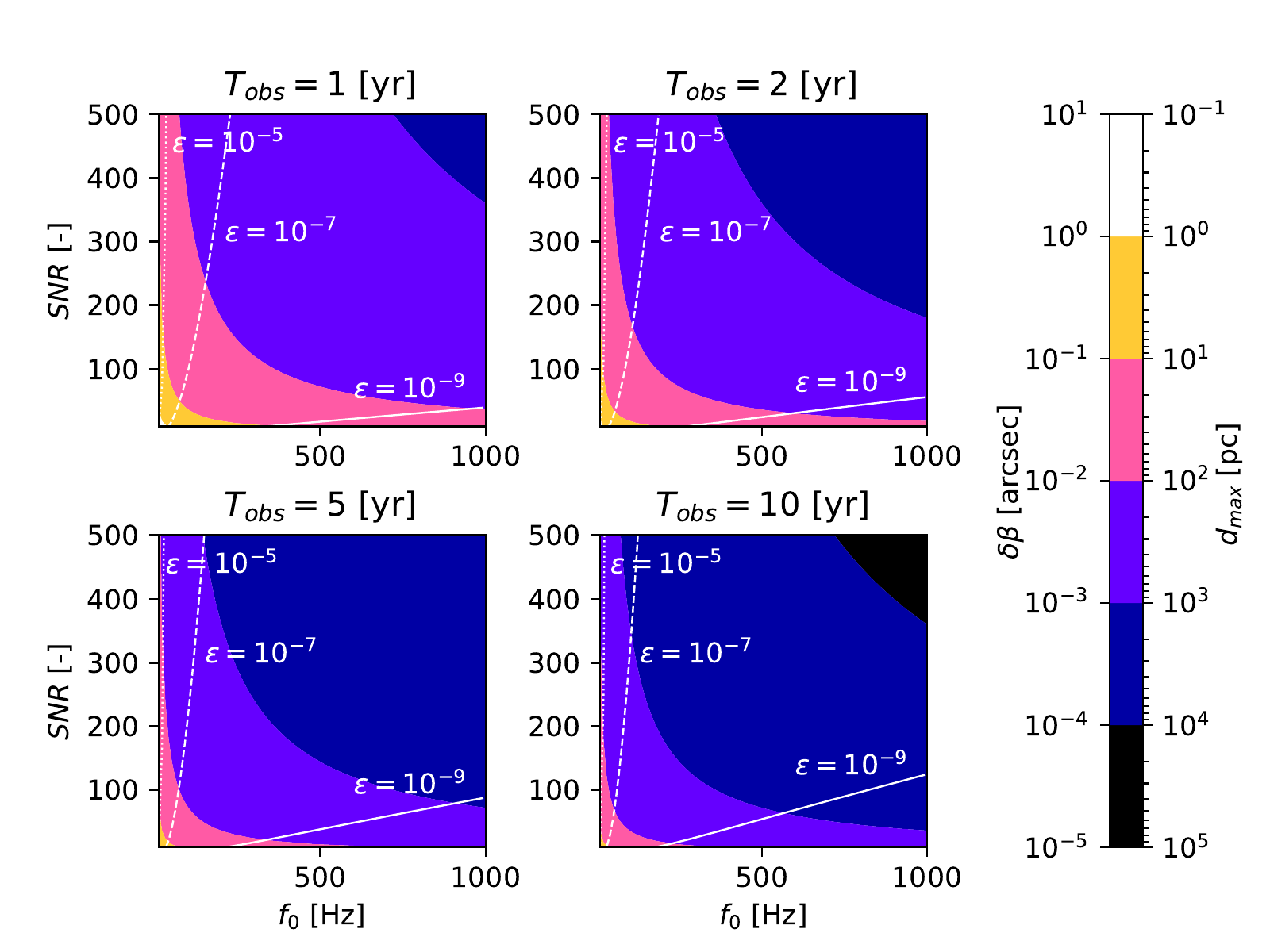}
    \caption{Same as for Fig.~\ref{fig:DB_aLIGO} but for the ET sensitivity.}
    \label{fig:DB_ET}
\end{figure}

In Figs.~\ref{fig:DB_aLIGO} and \ref{fig:DB_ET}, we present how the sky resolution depends on SNR, frequency $f_0$, and observation duration $T_{\rm obs}$ (to be consistent with \citet{Seto2005}, we present results for $\beta = \pi/4$). The twin axes of the colourbar represent the angular resolution $\delta \beta$ of the CGW, and the corresponding maximal distances $d_{\rm max}$ at which the parallax is resolvable.  The latter was calculated by setting $\delta \theta = \delta \beta$ and $d = d_{\rm max}$ in Eq.~\ref{eq:parallax_general}.  The white curves denote lines of constant SNR for stars of given ellipticity $\epsilon$, and an assumed distance to the source $d=130$ pc, which corresponds to the distance to the nearest known NS, namely RX J1856.5-3754 \citep{Drake2002}. We assume that all of the energy loss is due to the CGW emission, as is required for the analysis of \citet{Sieniawska2022} to apply (see \citet{Sieniawska2022}, and also the discussion in Sect.\  \ref{sec:discussion} below).  We calculated SNRs using the aLIGO sensitivity curve in Fig.~\ref{fig:DB_aLIGO} \citep{Abbott2020b} and the Einstein Telescope (ET) sensitivity curve in Fig.~\ref{fig:DB_ET} \citep{Sathyaprakash2012}. As expected, longer observation times, higher frequencies and larger SNRs allow parallax to be measured to larger distances.

\begin{figure}
	\centering
	\includegraphics[width=\columnwidth]{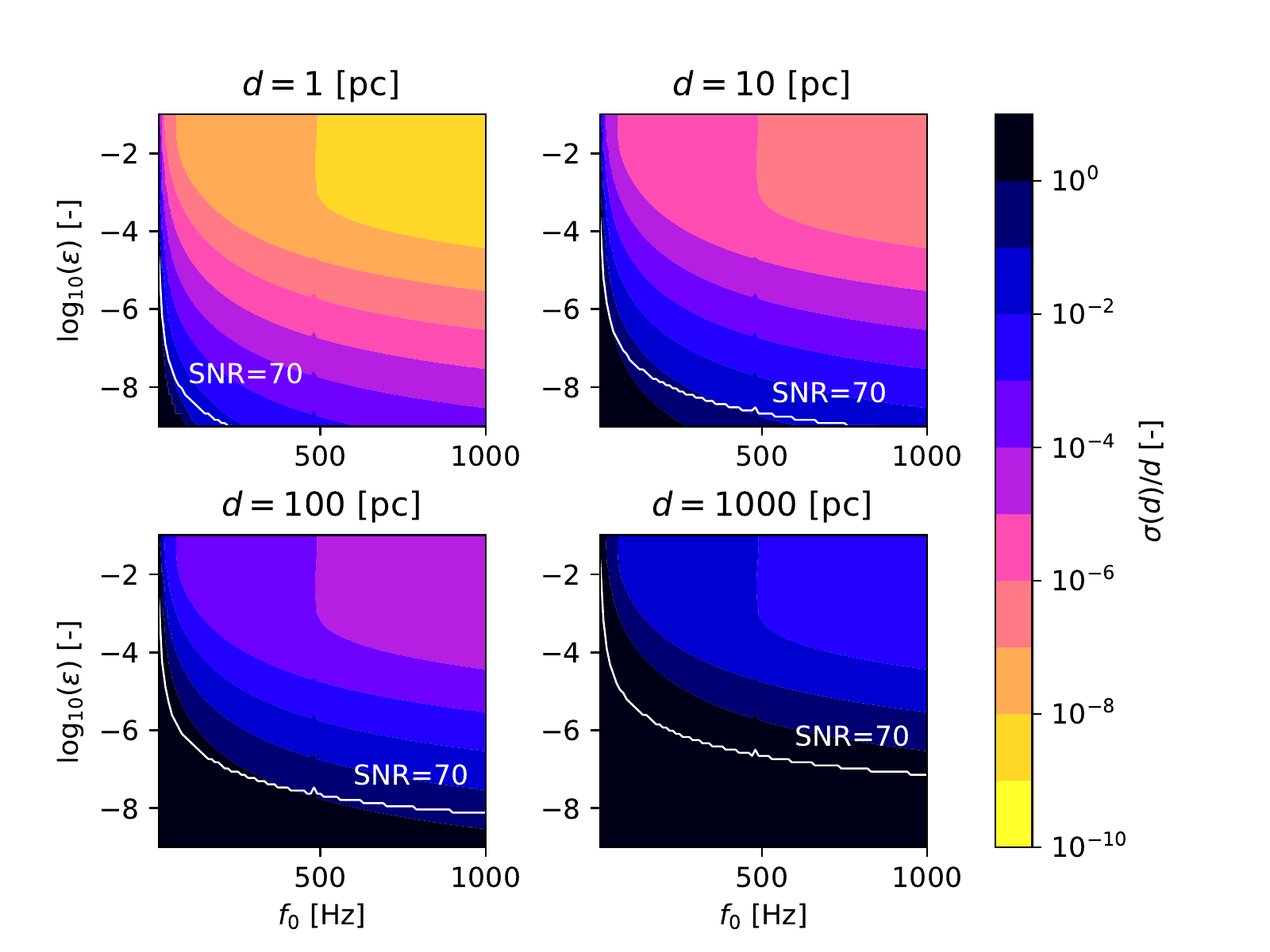}
    \caption{Relative error of the distance estimation with the parallax method for a broad range of possible parameters and assumed spin-down limit. aLIGO sensitivity curve is used and 2 years of observations are assumed.  White curves mark the threshold SNR=70.  }
    \label{fig:rel_err_MS2_aLIGO_epsilon}
\end{figure}

\begin{figure}
	\centering
	\includegraphics[width=\columnwidth]{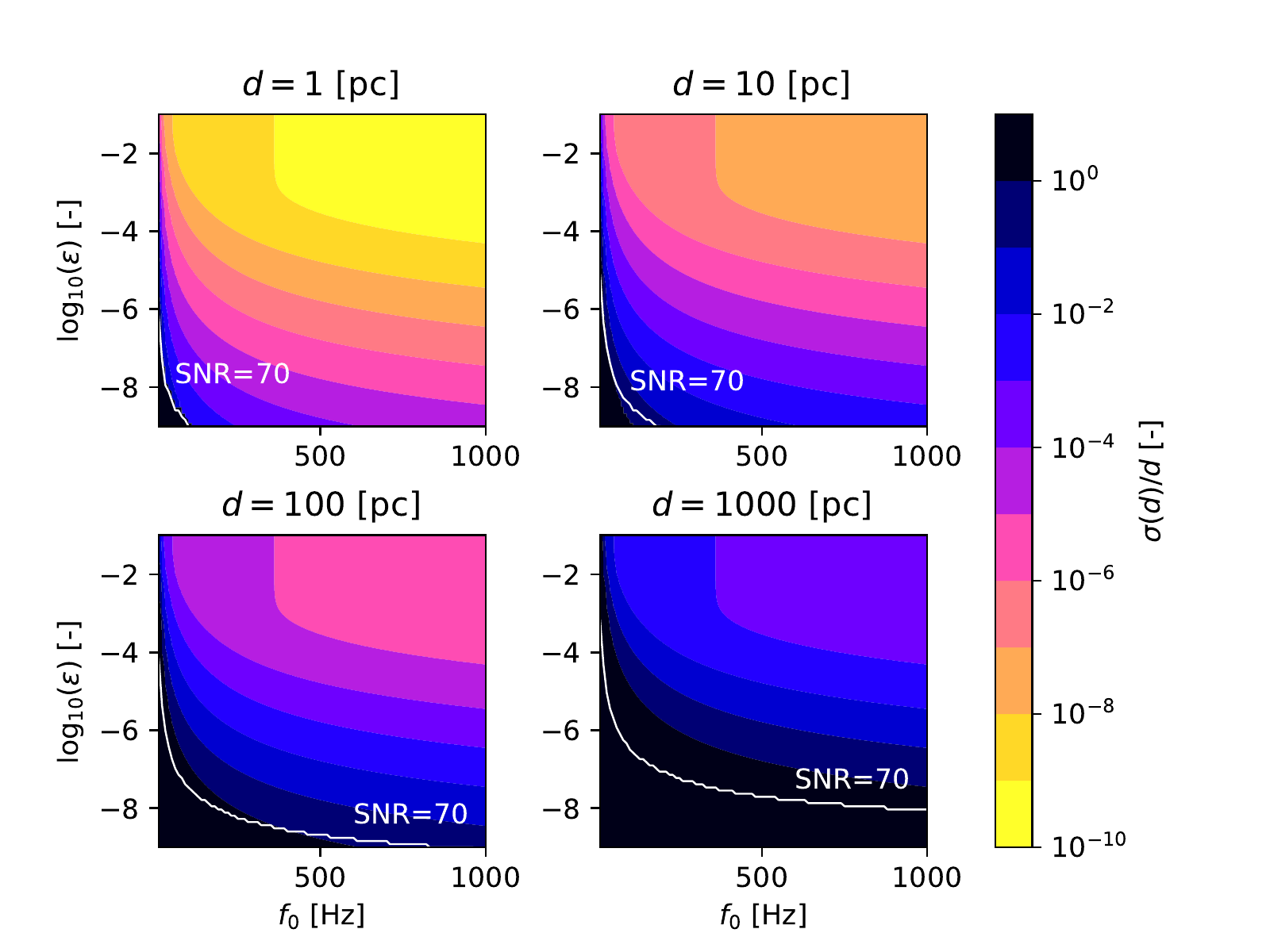}
    \caption{Same as Fig.~\ref{fig:rel_err_MS2_aLIGO_epsilon}, but for the ET sensitivity.}
    \label{fig:rel_err_MS2_ET_epsilon}
\end{figure}

The results of the relative error of the distance estimation with the parallax method, in terms of $\epsilon$ (for the assumed spin-down limit) are presented in Figs.~\ref{fig:rel_err_MS2_aLIGO_epsilon} and \ref{fig:rel_err_MS2_ET_epsilon}, using aLIGO and ET sensitivity curves, respectively. As expected, smaller distances, higher frequencies and larger ellipticities result in smaller relative errors in the distance estimation using the parallax method.

\begin{figure}
	\centering
	\includegraphics[width=\columnwidth]{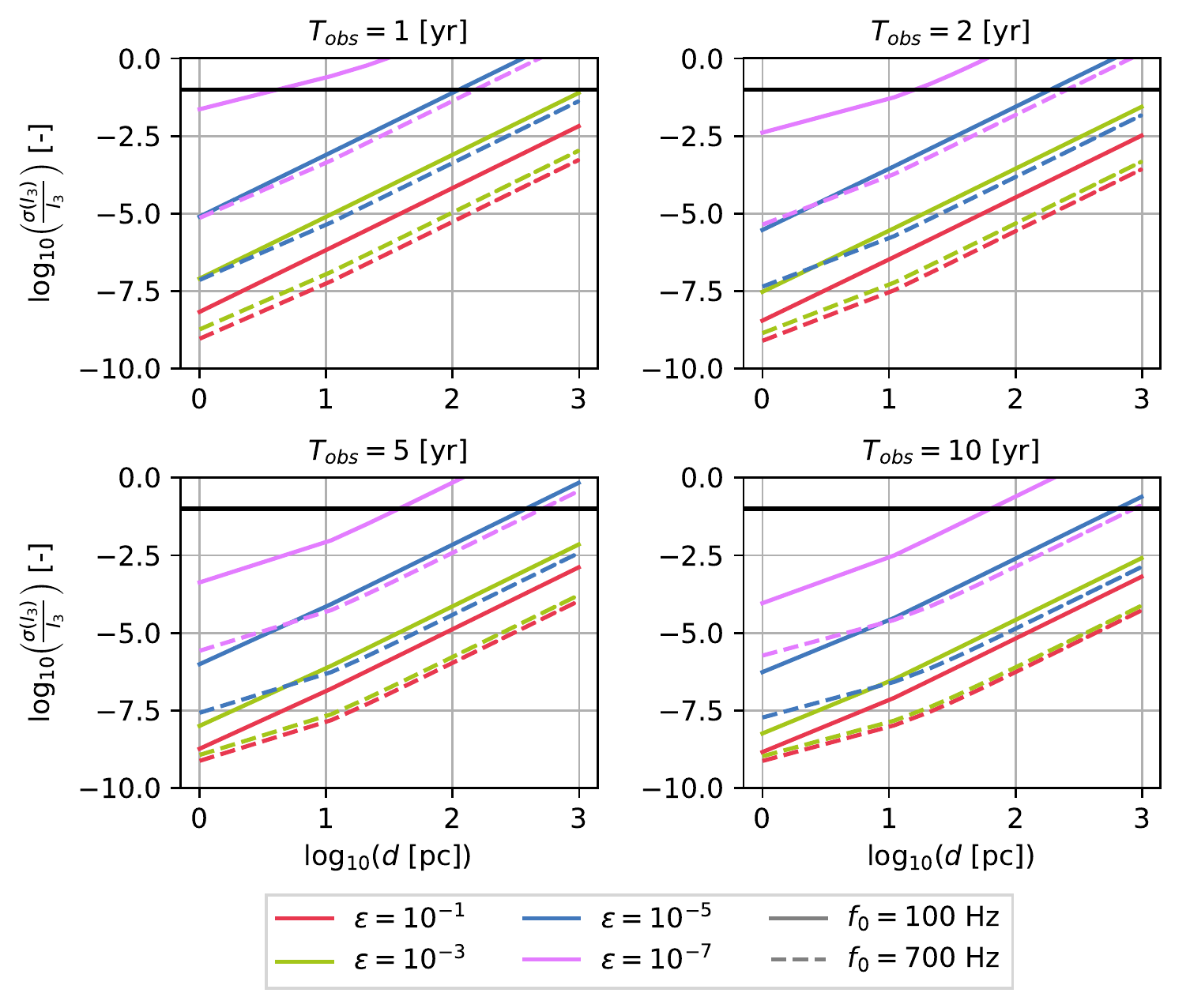}
    \caption{Relative estimation errors $\sigma(I_3)/I_3$ calculated from the combined measurements of $d$ and $d/\sqrt{I_3}$, for the ET sensitivity.  Black lines represent the 10\% relative uncertainty levels.}
    \label{fig:rel_err_I3}
\end{figure}

The relative errors of the $I_3$ estimation  as a function of source distance, as obtained with the combined measurements of $d$ and $d/\sqrt{I_3}$, are shown in Fig.~\ref{fig:rel_err_I3}, for several different values of observation time, ellipticity, and birth frequency. Here, we assumed $\sin\beta = \sin(\pi/6)$ and used the ET sensitivity curve. Results for aLIGO are approximately an order of magnitude worse but follow the same trend. We find that all solutions for $f_0>50$ Hz (so also all curves presented on the plot) reach higher SNRs than the typically used threshold SNR=70 \citep{Abbott2019, Abbott2022}, that indicates potentially detectable signals. The black lines on the plot denote $\sigma(I_3)/I_3 = 10\%$.

\begin{figure}
	\centering
	\includegraphics[width=\columnwidth]{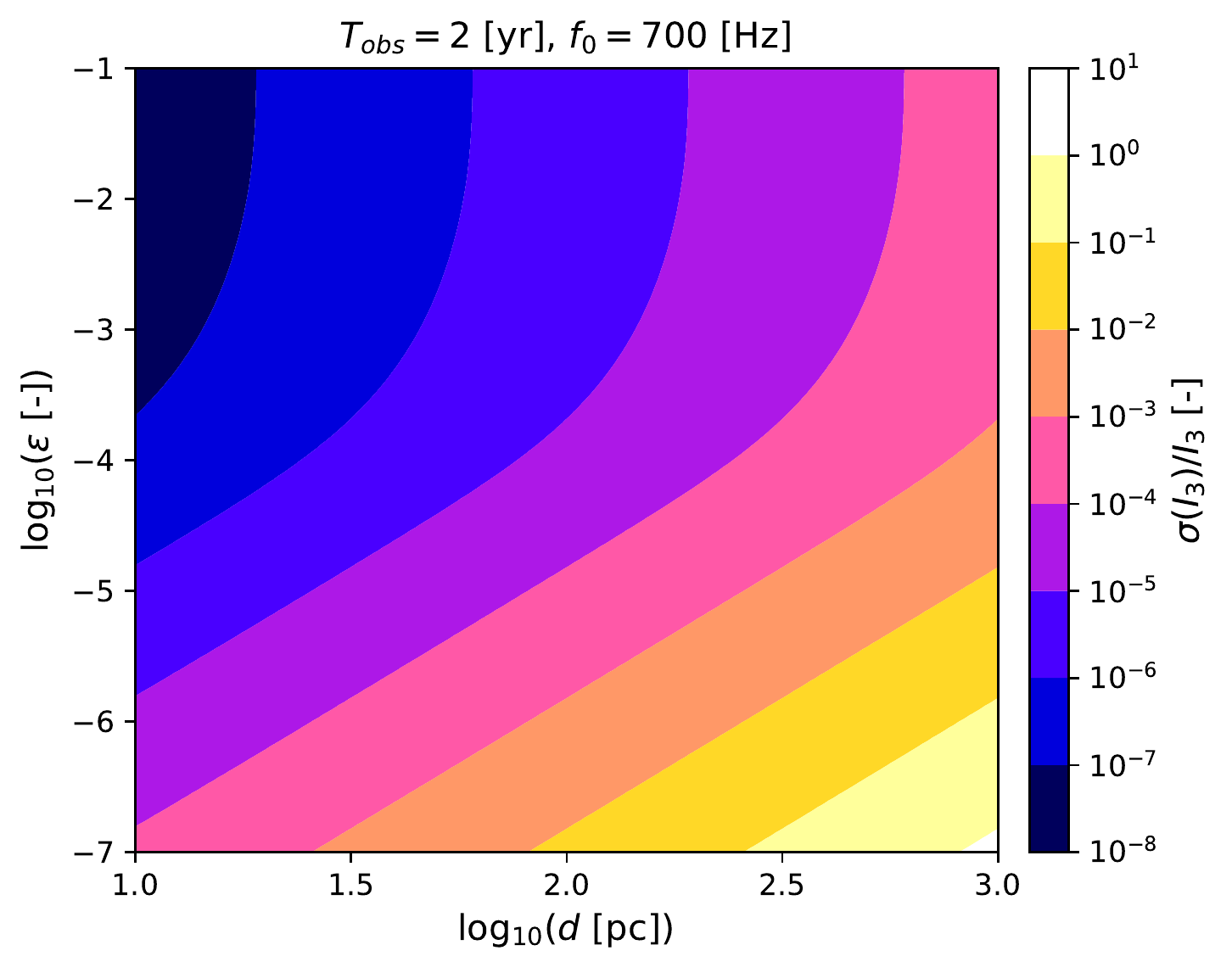}
    \caption{Relative estimation errors $\sigma(I_3)/I_3$ in terms of $d$ and $\epsilon$, for the values $f_0=700$ Hz and $T_{\rm obs} = 2$ years, for the assumed spin-down limit and ET sensitivity.}
    \label{fig:SD_I3}
\end{figure}

We can also illustrate the accuracy to which $I_3$ can be measured as a function of both $d$ and $\epsilon$, by fixing only the observation time and birth frequency, using a colour-scale to indicate the fractional error in $I_3$.  Such a plot is given in Fig.~\ref{fig:SD_I3}, where we fix $T_{\rm obs} = 2$ yr and $f_0 = 700$ Hz, and assume the ET sensitivity curve.  All solutions for the considered parameter space with  $d\lesssim 300$ pc satisfy the condition $\sigma(I_3)/I_3 < 10 \%$. We also checked that all solutions presented on Fig.~\ref{fig:SD_I3} satisfy the condition SNR>70.

\section{Discussion}
\label{sec:discussion}
The GW parallax method allows for a new opportunity to measure distances to CGW sources without extra, electromagnetic observations. In principle, it allows us to break the sorts of degeneracies inherent in CGW detections, that were discussed in \citet{Sieniawska2022}.

There is in fact a three-way degeneracy between distance $d$, ellipticity $\epsilon$, and moment of inertia $I_3$, as per Eq. \ref{eq:h0tr}.  The extent to which these degeneracies can be broken depends upon which features of the CGW signal that can be measured, as we now describe.

At a minimum, a CGW detection will provide a measurement of the signal amplitude $h_0$.  As is clear from Eq. \ref{eq:h0tr}, this would allow estimation only of the quantity $\epsilon I_3 / d = (I_2 - I_1) /d$, i.e. the ratio of the  (dimensional) asymmetry $(I_2 - I_1)$ to the distance.  Such a measurement in itself is of little use.  However, if the parallax effects, which were the main subject of this analysis, can be measured, one can extract $d$ and $I_2 - I_1$ individually.  

Measurements of distance are of interest in their own right, giving information on the distribution of NSs in the Galaxy.  In particular, it would be interesting to see if the spatial distribution of such `gravitationally selected' NSs differs from that of the known (electromagnetically selected) pulsar population.

Measurements of $I_2 - I_1$ are also useful, as the size of this quantity is directly proportional to the strength of the elastic and/or magnetic strains that sustain such an asymmetry.  There has been an extensive study of how the details of the microphysics of the crust limit such asymmetries, principally its breaking strain and shear modulus.  For examples see \citet{Ushomirsky2000, Owen2005, Haskell2007, Johnson-McDaniel2013, GAJ_21, GA_21, MH_22}.   There have also been a small number of studies of what evolutionary paths might produce such elastic strains in the first place, including \citet{setal_20, OJ_20, gc_22}.    For studies of how the star's magnetic field produces asymmetries, see for example \citet{BG_96, hetal_08, GJS_12}.  Actual measurements of $I_2 - I_1$ could then be interpreted in the context of this modelling, allowing us to set lower bounds on the sum of the maximum elastic strains that can be supported in solid phase(s), and the strength of the internal magnetic field.

If in addition spin-down measurements allow an estimate of the combination $d / \sqrt{I_3}$ using the methods described in \citet{Sieniawska2022}, then $I_3$ itself (as well as the dimensionless asymmetry $\epsilon$) can be measured. Note that this requires making the assumption of $100\%$ conversion of spin-down energy into GW energy, as the calculations in \citet{Sieniawska2022} are based upon equating the inferred rate of loss of rotational kinetic energy to the GW luminosity.  The quantity $I_3$ is a difficult one to measure, with the only direct observational constraints being a very weak upper bound, coming from the observed spin-down luminosity of the Crab pulsar \citep{bh_03}.    The most promising method for improving on this is currently believed to be from continued observation of spin-orbit coupling effects in the double pulsar \citep{ls_05}, which may give a $10\%$ error measurement of pulsar's moment of inertia in the near future \citep{ketal_21}.  The combination of parallax measurements with spin-down measurements described here offers an alternative direct measurement.

We showed that the relative error of the moment of inertia estimation $\sigma(I_3)/I_3$ could be smaller than $10\%$, especially in the case of small distances and relatively large ellipticities; see Fig.~\ref{fig:rel_err_I3}. The largest ellipticities we considered ($\epsilon>10^{-5}$) are certainly too large to be supported by neutron stars, and would require the existence of more exotic compact objects, such as quark stars.  \citep{Haskell2007, Knippel2009, GJS_12, Johnson-McDaniel2013}.  The existence of such exotic stars with large asymmetries would clearly make the analyses describe here easier to carry out.

A measurement of $I_3$ could also be useful to constrain the NS equation of state (EoS) \citep{bh_03, ls_05, getal_20}.  If $I_3$ is the only measured stellar parameter, all that can be done is rule out those EoS that are too soft to support such values; see e.g. Fig 2 in \citet{bh_03}.  If in addition the mass, radius, or some related combination (e.g. compactness) is known, then the EoS can be probed more quantitatively, with only those EoS that support the combination of the measured parameters being allowed.  Note, however, that the methodology of \citet{Sieniawska2022} can only be employed if the star's spin-down is, to a good approximation, driven purely by GW emission.  Such a star may not be as bright electromagnetically, making the determination of an additional parameter difficult.  See \citet{Sieniawska2022} for a discussion of this tension.

It may in fact be possible to carry out an analysis of the sort described here, even if there is a significant electromagnetic component to the spin-down, as described in \citet{lu_etal_22}.  Specifically, if one assumes that in the case of purely electromagnetic emission the braking index $n \equiv f \ddot f / \dot f^2$ is exactly equal to $3$ (as appropriate for electromagnetic dipole emission into a vacuum), then a braking index measurement  between $n=3$ and the CGW value of $n=5$ would allow one to divide the total spin-down torque into its electromagnetic and gravitational pieces and still solve for $I_3$ and $\epsilon$; see \citet{lu_etal_22} for details.  It would still be necessary for the distance to be known (e.g. by the parallax effect discussed here).  Note, however, that in the few cases where the braking index has been measured, the value has generally been \emph{less} than $3$, inconsistent with vacuum dipole emission \citep{lgs_12}.

Alternatively, a measurement of $I_3$ can be translated into constraints on other stellar parameters if one assumes the validity of various \emph{universal relations} between $I_3$ and other quantities, specifically between $I_3$ and tidal deformability \citep{lk_18} or between $I_3$ and compactness \citep{ls_05}.  But such calculations do not probe the EoS state itself.

\section{Conclusions}
\label{sec:conclusions}

The results presented here make quantitative the extent to which we can hope to break degeneracies between the distance $d$, birth frequency $f_0$, and ellipticity $\epsilon$ of a CGW source.  

Figures 2-5 give information on our ability to measure the distance to NSs via parallax.  To give a specific example from Figure 2, a source with birth frequency $400$ Hz, detected with an SNR of around $100$, would, in a one year observation with aLIGO, show signs of parallax if it lies within a distance $d_{\rm max} \approx 100$ pc from Earth, if our modelling assumptions are fulfilled. This can be interpreted as a best-case scenario for the GW parallax measurements, for the given parameters of the source. 

Figures 6-8 give information on our ability to additionally constrain the source's moment of inertia, if spin-down effects are also detectable.  To give a specific example from Figure 7, for a source with an ellipticity $10^{-7}$, birth frequency $700$ Hz and distance $d=300$ pc, in a one year observation with ET, the moment of inertia could be measured to an accuracy of $10\%$.  

 The extent to which parallax can actually be measured, and to which we can obtain measurements of the moment of inertia, depends upon the number of suitably close and rapidly spinning NSs that Nature provides.  Hopefully, CGW detections will soon be forthcoming, to help us answer this question.  In the meantime, Monte Carlo studies using realistic distributions of the NS birth rate, initial spin frequency, and spatial distribution, are needed to determine how likely we are to make such measurements.  It will also be necessary to refine existing CGW search methods to allow for parallax, ideally by explicitly incorporating the source's finite distance into the waveform model.

\section*{Acknowledgements}

This material is based upon work supported by NSF's LIGO Laboratory which is a major facility fully funded by the National Science Foundation.

The authors gratefully acknowledge useful discussions with David Keitel and Rodrigo Tenorio, and also the comments of the anonymous referee.

DIJ acknowledges support from the STFC via grant number ST/R00045X/1. A.L.M. is a beneficiary of a FSR Incoming Post-doctoral Fellowship. 

\section*{Data Availability}
No new data were generated or analysed in support of this research.



\bibliographystyle{mnras}
\bibliography{main} 







\bsp	
\label{lastpage}
\end{document}